\documentclass[a4paper, 5p]{elsarticle}

\usepackage[american]{babel} 
\usepackage{lineno}
\usepackage{amssymb}
\usepackage{amsmath}
\usepackage{graphicx}
\usepackage{dcolumn}
\usepackage{bm}
\usepackage[official]{eurosym} 
\usepackage{verbatim} 
\usepackage{booktabs} 
\usepackage{tabularx,array,booktabs}
\usepackage{threeparttable}
\usepackage{hyperref}
\usepackage{cleveref} 
\usepackage{url}
\usepackage[export]{adjustbox}

\usepackage{color}
\definecolor{red}{rgb}{1,0,0}

\usepackage{multirow}
\usepackage{float}
\usepackage{placeins}

\usepackage{pdflscape}
\usepackage{afterpage}
\usepackage{threeparttable}
\usepackage{capt-of}
\usepackage{booktabs}

\usepackage{pifont}
\newcommand{\cmark}{\ding{51}}%
\newcommand{\xmark}{\ding{55}}%

\newcolumntype{Y}{>{\raggedleft\arraybackslash}X} 
\newcolumntype{Z}{>{\raggedright\arraybackslash}X} 
\newcolumntype{A}{>{\centering\arraybackslash}X} 

\usepackage[table]{xcolor}

\begin{document}

\journal{Elsevier}

\begin{frontmatter}

\title{Site-dependent levelized cost assessment for fully renewable Power-to-Methane systems}

\author[fzj,ude]{Simon Morgenthaler\corref{cor}}
\ead{s.morgenthaler@fz-juelich.de}
\author[fzj]{Christopher Ball}
\ead{c.ball@fz-juelich.de}
\author[fzj]{Jan Koj}
\ead{j.koj@fz-juelich.de}
\author[fzj]{Wilhelm Kuckshinrichs}
\ead{w.kuckshinrichs@fz-juelich.de}
\author[fzj,col]{Dirk Witthaut}
\ead{d.witthaut@fz-juelich.de}

\address[fzj]{Forschungszentrum J\"ulich, Institute for Energy and Climate Research, Systems Analysis and Technology Evaluation, 52428 J\"ulich, Germany}
\address[ude]{University Duisburg-Essen, Faculty of Engineering, Energy Technology, Lotharstr. 1, 47048 Duisburg, Germany}
\address[col]{University of Cologne, Institute for Theoretical Physics, Z\"ulpicher Str. 77,
               50937 Cologne, Germany}
               
\cortext[cor]{Corresponding author}

\begin{abstract}
The generation of synthetic natural gas from renewable electricity enables long-term energy storage and provides clean fuels for transportation. In this article, we analyze fully renewable Power-to-Methane systems using a high-resolution energy system optimization model applied to two regions within Europe. The optimum system layout and operation depend on the availability of natural resources, which vary between locations and years. We find that much more wind than solar power is used, while the use of an intermediate battery electric storage system has little effects. The resulting levelized costs of methane vary between 0.24 and 0.30\,\euro/kWh and the economic optimal utilization rate between 63\,\% and 78\,\%. We further discuss how the economic competitiveness of Power-to-Methane systems can be improved by the technical developments and by the use of co-products, such as oxygen and curtailed electricity. A sensitivity analysis reveals that the interest rate has the highest influence on levelized costs, followed by the investment costs for wind and electrolyzer stack.
\end{abstract}


\end{frontmatter}


\sloppy

\section{Introduction}
\label{sec:introduction}

\subsection{Background to electricity based synthetic natural gas}
A comprehensive reduction of greenhouse gas emissions requires novel technological solutions for long-time storage and sector coupling \cite{RN47, RN64}. Renewable power sources enable the decarbonization of the electricity sector, but the temporal variability of wind and solar power remains a challenge. Power-to-Methane (PtM) could solve this problem, as it provides enough storage to cover long periods of low renewable generation. In addition, the large-scale availability of synthetic natural gas (SNG) can boost the decarbonization of transport, industry and heating and provide general flexibility options \cite{RN3}. Finally, PtM can also improve power grid operation and avoid congestion, as electric energy can be converted to SNG and transported via pipeline \cite{RN2}. Indeed, it has been estimated that under a scenario involving 85\,\% emissions cuts by 2050, one third of the produced power will have to be dedicated to producing synthetic natural gas or hydrogen \cite{RN1}.

However, if SNG is to be successful and useful, systems must be optimized and appropriate market conditions have to be in place. In order to meet the requirements for transformation of energy systems, electrolysis units should use power from renewable sources and, when necessary, resort to short-term storage units \cite{RN1}. Certain geographical locations have more favorable conditions for PtM, as the greater availability of wind and sun leads to higher capacity factors \cite{RN1}. If certain locations have comparative advantage in producing SNG, there is the possibility that it could be traded among countries using long distance pipelines. In this paper, we identify optimal locations in the EU for producing SNG and study the best configurations of SNG producing units at these sites.

\subsection{Market considerations for synthetic natural gas}
\subsubsection{Economic facets}
PtM technologies are currently not competitive with conventional alternatives, including steam reforming, because of the high capital costs of the electrolyzers \cite{RN48,RN5,RN6}. A variety of options to improve the competitiveness of PtM has been discussed in the literature. Technological improvements of the electrolyzers (e.g. higher maturity and increase of product yield) bringing down the capital cost \cite{RN6}, increasing the size of the unit to take advantage of economies of scale \cite{RN5}, and increasing the capacity factors \cite{RN6,RN7} would enhance the economic performance. 
Other market options for PtM involve marketing the co-products of PtM processes, namely oxygen and heat for district heating systems, contribution to frequency control services \cite{RN5} and potentially “higher value markets”, namely transportation fuels \cite{RN6}. 

If PtM options are considered desirable by the government and the public, its economic viability could be augmented through support mechanisms, such as premia for methane \cite{RN2,RN7}.

The source of electricity is an important consideration when analyzing the economics and system-value of PtM. Three possibilities for the electricity supply of PtM units include: (i) only drawing surplus electricity, (ii) market driven – mainly being supplied with surplus electricity, but also using electricity at times of high renewable production and (iii) they are directly connected to renewable power technologies, such as their own wind farm \cite{RN6}. In market driven operation (option ii), PtM units divert power from other consumers during periods of sparsity. Hence, this option suffers serious disadvantages in terms of system integration and flexibilization. Under the conditions of the study by \cite{RN6}, especially non-consideration of cost of surplus electricity, the surplus driven option is the most economically feasible as the underlying scenario assumes a high availability of surplus electricity. Hence, it would make economic sense to install PtM units in areas with high renewable penetration and low electricity demand \cite{RN6}. Here, we focus on the third option with a direct electricity supply, with an optional marketing of the local surplus.

It is important to understand the interdependencies between PtM units and other market actors in evaluating the economic case for its implementation, such as the implications of PtM technologies on the gas infrastructure, for instance. PtM technologies exacerbate the potential underuse of natural gas import and transport infrastructure, already experiencing a decline due to the increased penetration of renewables \cite{RN8}. Alternative ways of recovering investments in gas infrastructure, such as increasing the share of capital costs in the final cost of gas, may be necessary if SNG is expanded further \cite{RN8}. SNG producers must compete for surplus electricity with other users, namely Power-to-Heat suppliers and other electricity storage system operators \cite{RN6}. Hence, the availability of surplus electricity may be limited and potential PtM operators may have to either purchase electricity at higher prices or resort to using their own supply. This may adversely affect the economic viability of PtM. Similarly, SNG producers are dependent on a source of CO$_{2}$, which implies that other actors or the PtM operator must be prepared to capture this CO$_{2}$. Whilst there is currently a sufficient supply of CO$_{2}$ sold on the merchant market to meet demand, additional plants for CO$_{2}$ capture will be needed to meet increases in demand \cite{RN61}. Viable business models for CO$_{2}$ capture plants will have to exist in order to provide this further supply of CO$_{2}$. PtM can also be considered as CO$_{2}$-conversion technology \cite{RN61}. As the captured CO$_{2}$ is subsequently used as a feedstock for SNG production, PtM is also one possible carbon capture and utilization (CCU) technology. Generally, the availability of this CO$_{2}$ depends on the price of CO$_{2}$ emissions and thus on the regulatory framework \cite{RN8}. 

The economic feasibility of PtM systems will be determined by improvements in the technical performance of PtM systems, the advantages of their locations and the ability to market co-products. External factors will also affect the feasibility of expanding SNG, such as its compatibility with gas infrastructure and the availability of both surplus electricity and CO$_{2}$.

\subsubsection{Methodological background to levelized cost of synthetic natural gas}
\label{sec:methodological-background-to-levelized-cost-of-SNG}

In order to evaluate the economic viability of PtM projects, the levelized cost metric is used. Levelized costs are commonly used to compare different electricity generating technologies. In this context, levelized costs are defined as the price per unit of produced electricity, typically\,kWh, required for the investor in the plant to break even \cite{RN38}. They are essentially equal to the sum of annualized investment costs and the present value of the yearly average running costs of the plant divided by the average yearly electricity produced from the plant \cite{RN14}. The concept of levelized costs is readily generalized to other energy carriers, in particular SNG, to evaluate the economic competitiveness of different generating technologies. 
A consistent definition of levelized costs requires a specification of the system and its boundaries. Plant-level levelized costs are costs that sets the boundaries of the system at the point where the electricity is fed into the grid and where no additional costs in terms of infrastructure or system adaptation are considered \cite{RN38,RN52}. Correspondingly, most articles focus on the plant-based levelized cost of SNG, namely they do not generally consider the wider system value that may emerge from SNG and generally exclude considerations, such as grid connection fees and electricity taxes. This paper will also consider only this plant-based levelized cost of SNG. 
In this paper, the levelized cost of SNG are calculated as ratio of the annualized investment costs plus the fixed and variable operation and maintenance costs and the total amount of SNG following \cite{RN51}:

\begin{equation}
    \mathrm{
    LC_{SNG} = \frac{\sum{(af \times cost_{invest})} + \sum_{time}{(O\&M)_{var,fix}}}{\sum_{time}{SNG_{produced}}}
    }
\end{equation}

The investment costs are annualized using an annuity factor (af) which is derived from the interest rate (i) and lifetime (T) of a particular technology: 

\begin{equation}
    \mathrm{
    af = \frac{i \times (1+i)^{T}}{(1+i)^{T} - 1}
    }
\end{equation}

Under the annuity approach, the capital costs incurred by a project are, essentially, converted into yearly flows \cite{RN54} and added to the operational and maintenance costs. The cumulative costs are then compared to the annual flow of an energy carrier such as electricity or, in this case, synthetic natural gas \cite{RN55}. This annuity approach aims to facilitate the cost comparison of technologies with different lifetimes. Results are easily accessible but highly condensed and can thus give only an approximate assessment of the economic viability of a particular technology \cite{RN55,RN56}. The interest rate used is influential in determining the levelized cost and is derived from the Weighted Average Cost of Capital (WACC) \cite{RN15}, expressing the return required on money invested. Interest rates can differ depending on the type of investment; small investments in photovoltaics (PV), for instance, may have a relatively high share of equity financing whereas investments in large-scale energy projects are usually heavily debt-financed \cite{RN13}. The return on equity is higher than the return on debt, hence the distribution of financing makes a difference to the WACC \cite{RN16}.
For electricity utilities, private investors typically expect a return of between 7\,\% and 10\,\% on capital \cite{RN15,RN46}, but this can depend on the type of market in which utilities operate, the country in which they operate and the risk associated with the technology and electricity prices \cite{RN15}. In this study, an interest rate of 7.5\,\% is assumed – in 2017/2018, the average interest rate applied to German companies in the energy and natural resources sectors was estimated at 5.5\,\% \cite{RN17}. However, given the technological risks associated with SNG production, an interest rate of 7.5\,\% appears reasonable. The levelized cost of hydrogen, a major input in the SNG process, is heavily influenced by the cost of electricity for industrial customers \cite{RN18} and this is a crucial aspect in the analysis of the economics of hydrogen and synthetic methane.

\subsubsection{Input parameters for the levelized cost of synthetic natural gas}

Levelized costs of SNG strongly depend on the assumptions of technical and economic parameters and their future development. An overview of publications specifically discussing the levelized cost of SNG is given in Table~\ref{tab:literature_overview}. Input parameters featured in these studies will be covered here whereas a more detailed discussion of the technical configuration and technical scope of these publications will be given in section~\ref{sec:technical-and-economic-assumptions-researchgap}. 

Investment costs of SNG technologies are either given for a full PtM plant or separately for the electrolyzer and methanation reactor. Most studies provide cost estimates for the near future as well as an outlook to 2030 and 2050.

Estimates of the current investment costs of the electrolyzer vary quite strongly, ranging from 650\,\euro/kWh to 1,000\,\euro/kW (2020) \cite{RN25}, 1,000-2,000\,\euro/kW \cite{RN20} and 1,500\,\euro/kW \cite{RN26} and 2,000\,\euro/kW \cite{RN5}. As for future cost reductions for the electrolyzer, \cite{RN26} predict costs falling to 1,000\,\euro/kW by 2030 and 800\,\euro/kW by 2050, whereas \cite{RN20} and \cite{RN25} offer more optimistic estimates of reductions to 700\,\euro/kW and 1,200\,\euro/kW by 2030 and to 385\,\euro/kW and 660\,\euro/kW by 2050 and 500\,\euro/kW to 850\,\euro/kW (2030) and to 400 to 660\,\euro/kW by 2040 respectively. 

Investment costs of the methanation system are predicted to fall from 150\,\euro/kW \cite{RN26} and 160\,\euro/kW (2020) \cite{RN25} to 100\,\euro/kW (2030) and 70\,\euro/kW (2050) \cite{RN26} and 140\,\euro/kW (2030) to 125\,\euro/kW (2040) \cite{RN25}. As regards operating and maintenance costs, these are estimated at 3\,\% of the investment costs \cite{RN65}.
Electricity prices are commonly treated as external input variables. In many cases, they are derived from EEX spot data and sometimes tax-free electricity is factored in for certain scenarios \cite{RN5,RN26}. Other studies assume that PtM systems utilize surplus power, which is available at very low variable costs, such that only costs for grid access and transmission have to be covered \cite{RN28}. 

Certain papers discuss the costs for the supply of CO$_{2}$ in detail, factoring in the cost of technologies for CO$_{2}$ capture. The resulting cost estimates can diverge drastically even for similar technologies. In the case of direct air capture, for instance, estimates range from 200\,\euro/t \cite{RN5} to 1,000\,\euro/t CO$_{2}$ \cite{RN37}. The cost of CO$_{2}$ is omitted by \cite{RN25}, as they focus on PtM units in close proximity to large quantities of CO$_{2}$, such as distilleries. Co-products can serve to reduce the levelized cost of SNG, with the ETS certificates \cite{RN20}, O$_{2}$, heat and ancillary services, such as frequency control \cite{RN5,RN20}. The treatment of co-products will be discussed in detail in section~\ref{sec:treatment-of-co-products}. 

\subsubsection{Treatment of co-products}
\label{sec:treatment-of-co-products}

The production of SNG results in several co-products which can lead to additional revenue, displayed in the overview given in Table~\ref{tab:literature_overview}. The cost allocation is according to possible market prices of co-products. The most common co-product is O$_{2}$ and \cite{RN5} assumed a revenue from the sale of O$_{2}$ of 0.1 \$/kg, whereas \cite{RN23} did not give a direct figure, but suggested that this would be industrial oxygen which is sold at a lower price than medical oxygen. Furthermore, SNG production sites can offer ancillary services for power grid operation \cite{RN5,RN25}, heating \cite{RN5} and potentially even for CO$_{2}$ certificate trading \cite{RN23}. As regards ancillary services to the grid, \cite{RN25} indicated that transmission system operators would pay a fixed fee in return for these services and \cite{RN5} introduced the concept of levelized value of energy services. This metric captures the wider value of system services provided by PtM technologies, even including the avoided cost of produced or imported fossil fuels and gives a much deeper view of the possible economic benefits of SNG. Although the cost of SNG is far higher than that of natural gas, the additional value provided by SNG far exceeds that of natural gas, when these co-products are included in the analysis \cite{RN5}. Revenue from CO$_{2}$ certificate trading is small, but this is attributed to the low current price of certificates \cite{RN23}.

\subsection{Technical and economic assumptions and research gaps of previous literature}
\label{sec:technical-and-economic-assumptions-researchgap}

Remarkably, only very few prior studies have evaluated the levelized costs of SNG produced by PtM plants. This economic metric is extremely convenient to analyze system integration and optimization of new technologies and to assess their economic competitiveness. Levelized costs are widely used for the comparison of electricity generation technologies and has been generalized to a variety of other energy technologies as for instance hydrogen production \cite{RN9} or energy storage \cite{RN10}. First estimates of the levelized costs of SNG appeared around 2015 and only few other studies followed. An overview of ten research publications and their modelling scope is provided in Table~\ref{tab:literature_overview}.

\subsubsection{Spatial assumptions and research gaps}
With regard to the spatial scope, the studies featured in Table~\ref{tab:literature_overview} typically consider one individual country or the electricity grid mix of the European Union (EU) as a supranational region. None of the identified and analyzed studies contains a comparison of levelized costs for several countries. This finding coincided with an investigation conducted within the review study of \cite{RN49} that only identified one study with European focus beside several publications with national scopes. Assessments on subnational regions would provide in-depth insights. Regions with high renewable production and low electricity demand may provide optimal conditions for decentralized PtM units, as explicated for the German federal state of Ba\-den\--Würt\-tem\-berg in \cite{RN6}. However, if the gas is not to be consumed locally, but transported, a substantial expansion in gas transport infrastructure \cite{RN6} may be needed.

\subsubsection{Technical assumptions and research gaps}
There is a broad range of considered technical aspects assumptions of the studies considered in Table~\ref{tab:literature_overview}. The greatest transparency is given regarding the power rating of the system. Most studies focus on multi-megawatt systems. Only two articles consider small-scale systems: Parra et al point out results for $\leq$ 100\,kW \cite{RN5}, while McKenna et al study PtM systems of variable size between 0 and 200\,MW \cite{RN6}.

The main reason why studies on the levelized costs of SNG focus on large-scale PtM plants is probably the expected lower levelized costs compared to smaller units \cite{RN73}. However, recent projections revealed a broad potential for units with a nominal power below 100\,kW. In the case of Germany, a potential of around two thirds of all plants with a nominal electrolysis power below 100\,kW was calculated by \cite{RN24}. The use of small-scale PtM units shall in particular help to comply with the technical boundary conditions in the operation of distribution grids by integrating electricity from fluctuating renewable energy sources. 

The literature review, summarized in Table~\ref{tab:literature_overview}, demonstrates several research gaps in terms of the technical aspects of PtM. 

Whilst there is good coverage of the use of by-products, there is very limited discussion of battery storage – which is crucial for self-sufficient systems. As regards the exogenous factors affecting the performance of PtM units, studies appear to have neglected the capacity factors and the influence of natural climate variability (see \cite{RN57,RN59} for discussion on natural climate variability). 

These technical research gaps have an influence on economic parameters underpinning the levelized cost metric.

\subsubsection{Economic assumptions and research gap}

Levelized costs provide a convenient metric for techno-economic assessments, but the treatment of different contributions is neither straightforward nor consistent in previous literature. In particular, studies differ in the inclusion of different cost components, the source of the electricity used to power the unit and the inclusion of taxes and depreciation.

Cost components concerning the electrolysis and methanation technology are part of all considered publications. While methanation focuses on thermo-chemical concepts and rarely considers biological methanation, a variety of electrolysis technologies are assessed. The three main variants considered are: polymere electrolyte membrane (PEM) electrolysis, alkaline water electrolysis (AWE) and solid oxide electrolysis cells (SOEC). Cost components concerning infrastructures (e.g. pipelines) are not commonly used and considered only in four out of ten publications.

O\&M costs are clearly stated in most publications. Only one study does not clearly specify O\&M costs. Furthermore, as the life cycle of the involved PtM technology is limited, some studies involve costs for replacement or disposal of system components.

Electricity costs are substantial components of the levelized costs as the hydrogen production is based on electrolysis. The handling of electricity costs varies between the considered studies but spot market / wholesale electricity prices have been considered most frequently. For a case study about electrolysis application in Baden-Württemberg in 2040 \cite{RN6} consider different possibilities for the sourcing of electricity. The required utilization rate is lower if all of the power comes from using surplus electricity than if this power is purchased on the market (grid electricity) or comes from direct coupling (i.e. a PtM unit installing its own renewable power sources) \cite{RN6}. 

Ideally, from an environmental perspective, PtM systems should preferably be operated with renewable energies such as wind energy plants and not by fossil energy sources \cite{RN62}. However, for an economically efficient running of the system, it is best if the electrolysis unit can draw on power from “short-term” storage devices to ensure that the units have sufficient full load hours to be economically feasible \cite{RN1}.

\afterpage{
    \clearpage
    \thispagestyle{empty}
    \begin{landscape}
        \centering
        \begin{table}[tb]
    \rowcolors{2}{gray!25}{white}
    \begin{tabularx}{9.5in}{ZAAAAAAAAA}
    \rowcolor{gray!50}
    \toprule
         Author & Spatial scope & System power & \multicolumn{3}{c}{System configurations (technical scope)} & \multicolumn{2}{c}{System analytic assessments} & \multicolumn{2}{c}{Economic assessment} \\
     \midrule
          & & & Type of electricity & Coupling with battery storage & Byproduct(s) & Influence of individual weather years & Analysis of different local wind and PV capacity factors & Infrastructure costs & Inclusion of fiscal details, incl. CO$_{2}$ certificates \\
      \midrule
      Schiebahn et al. \cite{RN3} & DE & n.s. & Wind? & \xmark & \xmark & \xmark & \xmark & \cmark & \xmark \\
      Thomas et al. \cite{RN20} & BE & 15\,MW & Grid & \xmark & \cmark & \xmark & \xmark & \cmark & \cmark \\
      Parra et al. \cite{RN5} & CH & 25\,kW to 1000\,MW & Grid & \xmark & \cmark & \xmark & \xmark & \xmark & \cmark \\
      Gutiérrez-Martín et al. \cite{RN28} & ES & 50\,MW & Grid & \xmark & \cmark & \xmark & \xmark & \xmark & \xmark \\
      De Bucy et al. \cite{RN26} & n.s. & 10\,MW & Grid & \xmark & \xmark & \xmark & \xmark & \cmark & \xmark \\
      Glenk et al. \cite{RN6} & DE & 0 to 200\,MW & RES-based grid & \xmark & \xmark & \xmark & (\xmark) & \xmark & \cmark \\
      McDonagh et al. \cite{RN25} & IE & 10\,MW & Grid & \xmark & \cmark & \xmark & \xmark & \xmark & \xmark \\
      De Vita et al. \cite{RN21} & EU Mix & n.s. & ? Grid & \xmark & \xmark & \xmark & \xmark & \cmark & \xmark \\
      Salomone et al. \cite{RN22} & DE & 1\,MW to 20\,MW & RES-based grid & \cmark & \xmark &\xmark &\xmark &\xmark & \xmark \\
      Guilera et al. \cite{RN23} & DE, SE, ES, PY, IN, Ont., Calif. & 10\,MW & Grid & \xmark & \cmark & \xmark & \xmark & \xmark & \cmark \\
  \bottomrule
    \end{tabularx}
        \captionof{table}{Overview of literature on levelized cost of synthetic natural gas.}
        \label{tab:literature_overview}
        \end{table}
    \end{landscape}
    \clearpage
}

For PtM to thrive, it must be in stakeholders’ economic interest and this is partly dependent on taxes, grid costs and incentives which are not typically considered in the literature. On the level of the market actors, subsidies to support long-term storage technologies, such as tax credits for investment in electrolyzers may be important \cite{RN66}. On a societal level, the provision of subsidies and the adaptation of infrastructure to accommodate PtM implies a social cost which relies on the consent of the public. These wider issues related to actors’ perspective in relation to PtM will be considered in more depth in further research.

Summarizing this short literature survey of existing studies revealed a common basis of understanding around the levelized costs of SNG. Furthermore, the studies delivered a valuable contribution to the establishment of levelized costs as economic metric for PtM assessments. However, diverging scope and consideration of different cost components have, hitherto, led to diverse results and making the comparability of former publications difficult.

\section{Model system configuration}
\label{sec:model-system-configuration}

\subsection{Description of island system}
\begin{figure}[tb]
    \centering
    \includegraphics[width=\columnwidth]{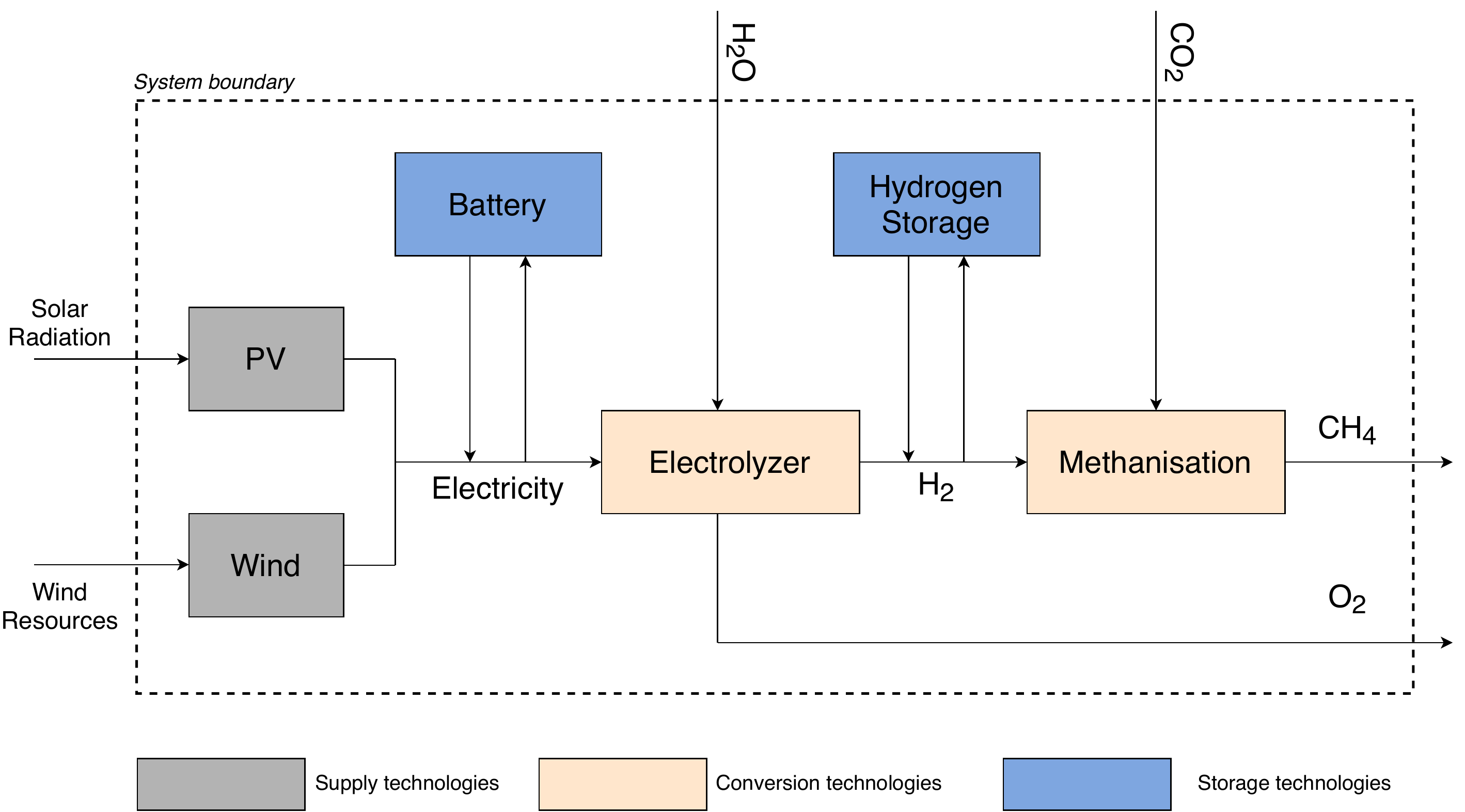}
    \caption{Considered process chain.}
    \label{fig:processchain}
\end{figure}

Different concepts exist for PtM plants as summarized in the previous section. In this article we consider a plant coupled to its own renewable electricity supply technologies (island system). An island system operates fully self-sufficient, has no connection to the grid and can thus neither draw nor feed electricity to the grid. This ensures a product with a minimal carbon dioxide footprint, as the only electricity sources are renewable. The process chain is shown in Figure~\ref{fig:processchain}. SNG consists of methane (CH$_4$) and is produced from hydrogen (H$_{2}$) and carbon dioxide (CO$_{2}$) in a methanation reactor. Hydrogen is generated in a PEM electrolyzer of electricity and water. We assume a small scale system with a fixed electrolyzer capacity of 100\,kW$_{el}$. As a co-product oxygen is produced which may be utilized as a by-product and sold commercially, leading to additional revenue. For this analysis, the electricity can be provided by either wind, PV or a combination. A battery storage unit is included as an intermediate electricity storage device to overcome shortfalls of supply. The hydrogen storage unit provides a buffer between the electrolyzer and methanisation reactor. Costs for water are minimal, so that the influence on economic performance of European PtM plants is negligible. 

\begin{figure}[tb]
    \centering
    \includegraphics[width=\columnwidth]{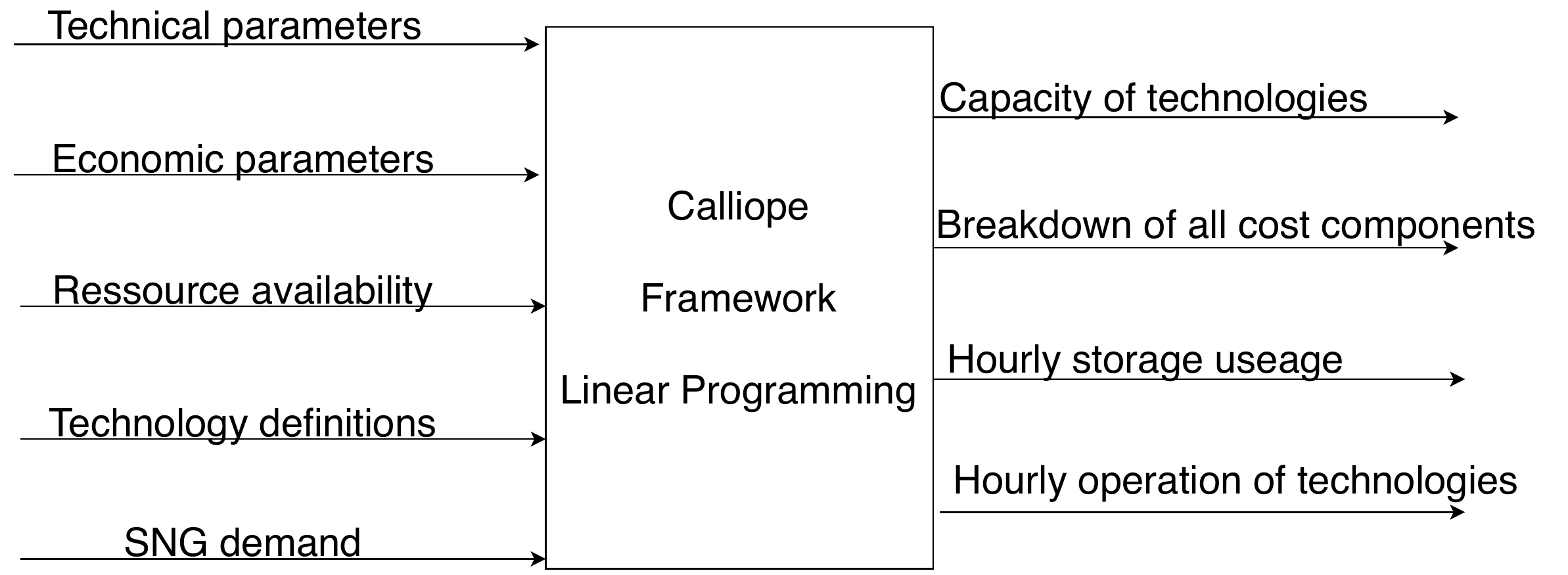}
    \caption{In- and outputs of our Calliope SNG model.}
    \label{fig:input-output-calliope}
\end{figure}

Calliope is an open-source energy system modeling framework written in Python and used in numerous energy system modelling studies (e.g. \cite{RN51,RN33,RN31,RN32,RN30}). The structure, functioning and mathematical aspects of this modeling framework are described in detail in the documentation \cite{RN33}. Calliope uses all model input data (cf. Figure~\ref{fig:input-output-calliope}) and creates a mathematical description of the energy system, which consists of linear equations. Calliope then optimizes the model by minimizing the total system costs, resulting in the cost-optimal capacity and operation of all technologies under consideration of all set constraints and equations. Storage size and usage is included in the optimization process. A simplified overview of in- and outputs of a Calliope model are shown in Figure~\ref{fig:input-output-calliope}. Within the definition of technologies the modeler is able to set constraints which must be respected at all times. Possible constraints can be the minimum load of a technology, a maximum load change per hour or a charge rate. Our model has an hourly time resolution. The objective function is minimizing the levelized costs of SNG over the entire year modeled.

The island system under consideration use renewable power from its own local sources. Data on the local availability of wind and solar is obtained from the platform Renewables.ninja, which offers bias-corrected capacity factors for wind and PV for Europe \cite{RN36,RN35}. The capacity factor is the ratio of the actual power output and the maximum possible power output of a renewable source in a given period of time. Capacity factor time series are available for any location in Europe with a spatial resolution on NUTS2 \cite{RN37} for the years 1980 – 2016. Capacity factor time series have the same temporal resolution as the model, such that no coarse-graining or data aggregation is applied. Thus it is possible to analyze the interaction between wind, PV and battery with the electrolyzer and the interaction between the electrolyzer and hydrogen storage with the methanisation reactor on an hourly basis.

\subsection{Selection of regions}
In this study we analyze PtM for two regions in Europe with different climatic conditions and thus different operational conditions for renewable power generation. Furthermore, the two regions are connected to the European gas grid with a capacity higher than 300 GW. Such a high interconnection capacity allows to connect PtM plants and to transport SNG in large volumes in the future. One region is in the north of the Netherlands (NL32: ‘Noord-Holland’), the other region is in Spain (ES61: ‘Andalucía’). 

Calliope is designed to perform multiple runs of the same model with different data. Thus, it is possible to analyze the influence of different weather conditions, using for instance different locations during the same weather year or different weather years at the same location. For this work the available capacity factors for wind and PV of 37 years for both regions in Europe (1980 – 2016) were collected. With this data each model run is performed for one year consisting of 8760 or 8784 hours. 

\subsection{Operating parameters for modeled Power-to-Methane plant}
The optimum layout of a PtM plant and the resulting costs of SNG strongly depend on techno-economic parameters such as investment costs, lifetimes or efficiencies. The values of these input parameters used in this study are summarized in Table~\ref{tab:model-parameters}.

\begin{table*}[tb]
    \centering
    \rowcolors{2}{gray!25}{white}
    \begin{threeparttable}
    \begin{tabularx}{\textwidth}{ZYYYYYYY}
    \rowcolor{gray!50}
    \toprule
    & Investment cost [\euro/kW] & Investment cost storage [\euro/kW] & Variable O\&M & Fix O\&M [\euro/kW] & Lifetime & Efficiency & Min. Usage of capacity [\%] \\
    \midrule
    Battery & 140 & 109 & 1.94 [\euro/MWh] & 6.4 & 15 & 0.954 & \\

    CO$_{2}$ Supply & & & 0.1 [\euro/kg] & 25 & & & \\

    Electrolyzer system\tnote{1} & 1,000 & & & & 20 & 1 & \\

    Electrolyzer stack\tnote{1} & 1,000 & & & & 6 & 0.6 & \\

    Hydrogen storage\tnote{2} & & 23.4 & & & 25 & 1 & \\

    Methanation\tnote{3} & 635.8 & & & & 20 & 0.85 & 40 \\

    Photovoltaic & 1,074 & & & 25 & & \\

    Wind onshore & 1,312.5 & & & 25 & & & \\

    Oxygen\tnote{4} & & & 0/-1/-10 [Cent/kg] & & & & \\

    Curtailment\tnote{4} & & & 0/-1/-5 [Cent/kWh] & & & & \\
    \bottomrule
    \end{tabularx}
    \begin{tablenotes}
    \item[1] Based on \cite{RN70}.
    \item[2] Based on \cite{RN68,RN69}.
    \item[3] Based on \cite{RN73,RN74}.
    \item[4] Revenue for oxygen and curtailed power at different magnitudes is considered in section~\ref{sec:use-of-coproducts} only.
    \end{tablenotes}
    \caption{Techno-economic model input data (from \cite{RN51} if not stated otherwise).}
    \label{tab:model-parameters}
    \end{threeparttable}
\end{table*}

The electrolyzer is divided into two parts, the stack component which has a lower lifetime and the rest of the plant with a higher lifetime. Distinguishing between two parts enables a more detailed analysis regarding the overall cost contribution, as well as a more detailed sensitivity study, e.g. influence of stack lifetime improvements. The overall efficiency is assumed to be 60\,\% whilst the total investment costs for the electrolyzer system are 2000 \euro{}/kW$_{el,in}$.

The methanisation reactor has a minimum usage of capacity, which means that it can be operated flexibly above the value but must not fall below the value.

\section{Research aims \& objectives}
This study aims on closing gaps on former levelized costs research on PtM by presenting a detailed production site analysis and focusing on small-scale decentral units. For this spatial differentiated analysis different local wind and solar power characteristics within two regions in Europe are evaluated. As the generation profile for wind and PV vary on many timescales (short-term, mid-term and long-term \cite{RN57}) this study includes 37 different weather years from 1980 – 2016.

A further goal of this assessment is a consistent technological modelling and economic assessment of PtM plants. Additionally, the study should not end at a tech\-no\--eco\-nomic boundary, as given for former studies. Rather, this assessment should additionally include system analytical considerations to enable broader insights in interactions on energy system level.

It can be assumed that a high utilization rate is economically favorable for PtM plants, thus a grid connection and a constant electricity price seem beneficial. However, within an island system electricity cannot be evenly supplied over all time steps. This leads to the question of whether part load operation can be economically favorable for island systems running on intermittent electricity supply technologies. 

During part load operation, the SNG demand is lower than the maximum capacity of the plant allowing the optimization model to flexibly adjust the output per hour. The loading level is commonly given in terms of full load hours per year such that the ratio of full load hours divided by 8760 hours (8784 hours in leap years) equals the ratio of demand and maximum capacity. Higher load levels (more full load hours) correspond to a better utilization of the electrolyzer and methanisation reactor, but require additional storage and generation capacity to provide the necessary electricity. Hence we are led to the questions: which load leads to the best economic performance and how will it vary between the years? Will the levelized cost of SNG vary strongly between the years? How will storage sizes and usage vary, if they are needed at all?

The current modeling framework computes the optimum system layout and operation for a given predefined demand of SNG. Hence, the optimum loading level is not computed automatically, but must be determined in an additional step. To this end, the optimization is run repeatedly for each region and weather year, varying the loading level in 50 steps between 5000 full load hours and 7800 full load hours. Finally, we choose the loading level with the best economic performance for each region and weather year. Table~\ref{tab:flh_ur_sngamount} gives an overview of the interrelation between full load hours, utilization rate and the resulting SNG amount.

\begin{table}[tb]
    \centering
    \rowcolors{2}{gray!25}{white}

    \begin{tabularx}{\columnwidth}{ZYY}
    \rowcolor{gray!50}
    \toprule
    Full load hours [h] & Utilization rate [\%] & SNG annual sum [t] \\
    \midrule
    8760 & 100 & 32.2 \\
    7800 & 89 & 28.6 \\
    5000 & 57.1 & 18.4 \\
    0 & 0 & 0 \\
    \bottomrule
    \end{tabularx}
    \caption{Full load hours, utilization rate and resulting SNG amount of the investigated system.}
    \label{tab:flh_ur_sngamount}
\end{table}

Additionally, we execute an analysis of the influence of the battery storage on the economic system performance and the economically favorable full load hours. For this we perform a second run-cycle with all the specifications listed above but without the battery storage option to compare results. 

System operation without storage is possible only if the electricity demand is fully flexible. Carmo et al. suggest a lower bound of the dynamic range for a PEM electrolyzer of 0\,\% – 10\,\% \cite{RN40}. In this article the lower value is chosen (0\,\%) to cope with the comparison of no battery option. A value greater zero of the electricity consuming technology (PEM) combined with no electricity storage option can lead to infeasibilities (time steps where no wind and PV supply exists in combination of a constraint that supply must be available). 

Considering island systems with no connection to the grid leads to potential curtailment, because any surplus electricity cannot be exported out of the system boundaries. Curtailment appears in time steps when the electricity generation technologies produce more electricity than the electrolyzer and battery can utilize. Curtailment is highly undesirable, as it corresponds to a loss of electricity that could potentially be used for other uses. How much curtailment will appear and does it fluctuate strongly across the different years?

To check which parameters have the highest influence on modeling results, a sensitivity analysis for both regions for the year 2016 is performed. The analysis includes the investment cost and lifetime of all components as well as carbon dioxide supply costs. To complete the picture we vary the economic value of oxygen, which may be sold, thus lowering the total system costs.

\section{Results}
In this section we discuss the optimum system layout and operation of fully renewable SNG production using the modeling framework introduced above. We first consider the operation in section~\ref{sec:optimum-system-operation}, illustrating its variability and the role of storage and curtailment. The optimum system layout crucially depends on the characteristics of renewable power generation as discussed in section~\ref{sec:effects-of-spatial-temporal-choices}. The economic competitiveness of PtM is then analyzed in section~\ref{sec:contribution-of-different-technologies-to-costs}, where we evaluate the cost structure of PtM plants and the resulting levelized costs of SNG. We finally discuss possible routes to improve the economic competitiveness of PtM. We analyze the use of co-products (section~\ref{sec:use-of-coproducts}), the importance of storage units (section~\ref{sec:importance-of-battery-storage}) and discuss the impact of future technical developments via a sensitivity analysis (section~\ref{sec:sensitivity-analysis}).

%

\subsection{Optimum system operation}
\label{sec:optimum-system-operation}
\begin{figure}[tb]
    \centering
    \includegraphics[width=\columnwidth]{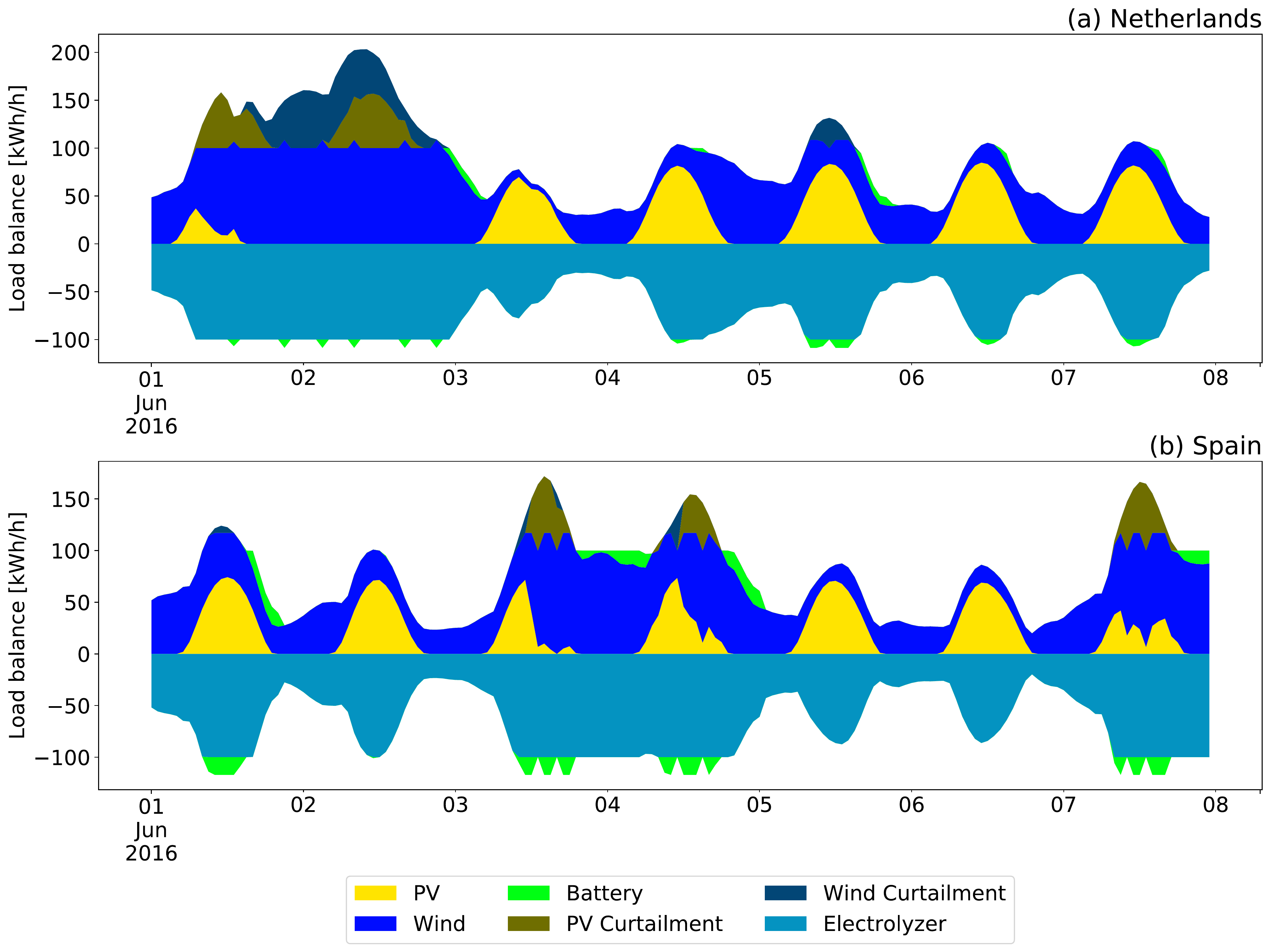}
    \caption{Load balance in (a) Netherlands (b) Spain for the first week of June in 2016 (hourly resolution).}
    \label{fig:load_balance}
\end{figure}

The operation of PtM plants is optimized at an hourly resolution. Figure~\ref{fig:load_balance} shows electricity generation and consumption by the different system components for one week. Renewable power generation varies strongly during this week. We clearly observe the daily cycle of solar power generation, but also days with vanishing generation. The operation of the electrolyzer mostly follows the power generation. On days with low wind power generation, one clearly discovers the daily cycle of solar power. The battery is mostly used for balancing on a daily time scale: It is charged during noon and discharged in the evening to enhance the daily utilization. Curtailment of renewable power generation occurs frequently, in particular during windy and sunny hours.

\subsection{Effects of spatial and temporal choices on peak capacities and size of storage devices}
\label{sec:effects-of-spatial-temporal-choices}

\begin{figure}[tb]
    \centering
    \includegraphics[width=\columnwidth]{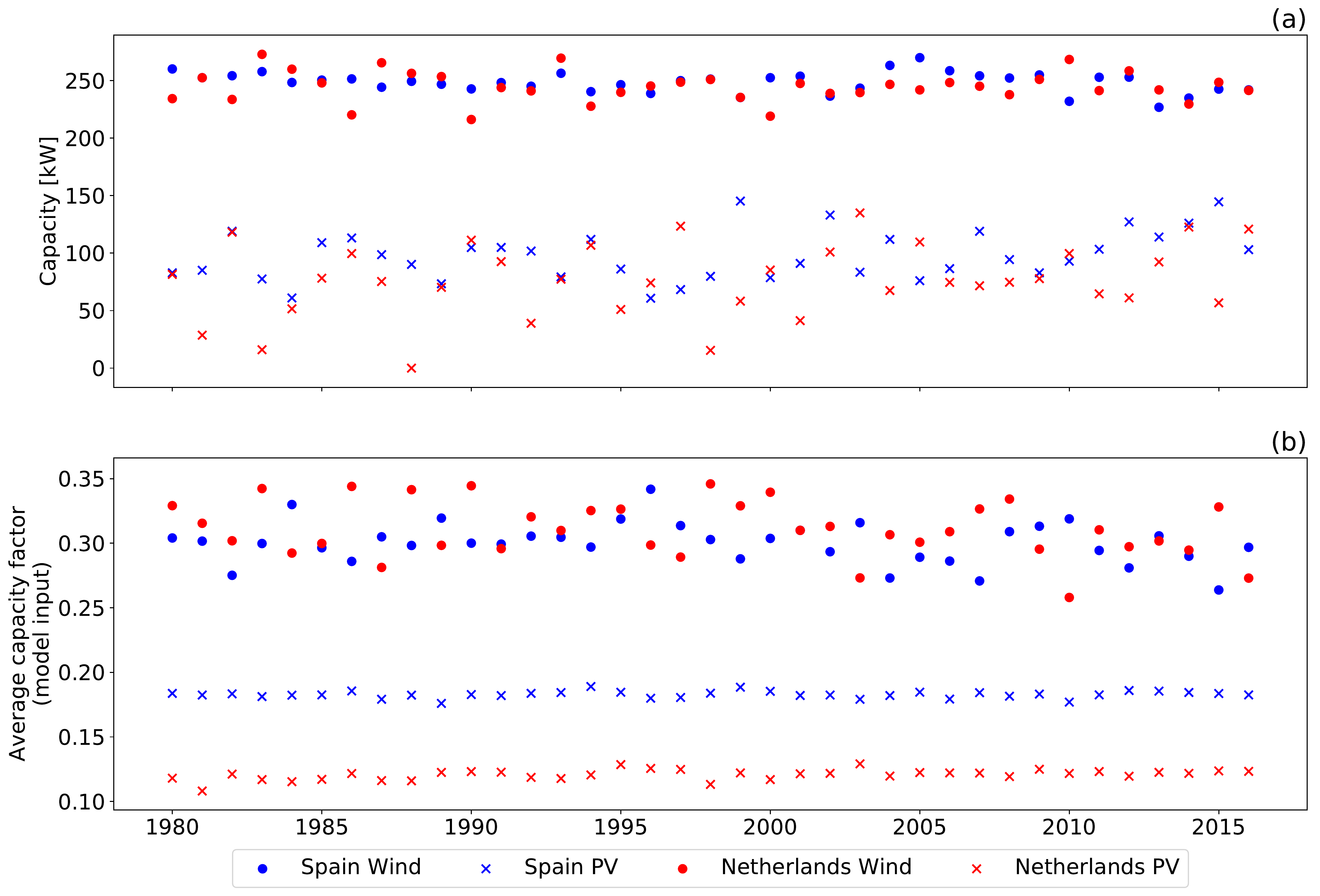}
    \caption{(a) Cost-optimal capacities of wind and PV for both regions and all years and (b) model input data (average capacity factors for wind and PV).}
    \label{fig:04_capacities_cf}
\end{figure}

Different conditions lead to different utilization of wind and solar power as shown in Figure~\ref{fig:04_capacities_cf} and consequently to differences in the optimum system layout. The region ES61 in Spain is located around 37$^{\circ}$ latitude, while the region NL32 is located around 53$^{\circ}$ latitude, and thus has a significantly higher solar irradiation. As a consequence, average capacity factors are higher by a factor of approximately 1.5. In contrast, little differences are observed in terms of the wind power resources. The difference in capacity factors in Spain and the Netherlands is smaller than the inter-annual variability. In both regions the inter-annual variability is stronger for wind than for solar power.

The cost-optimum system layout contains significantly more wind than solar power, for both regions and all years (Figure~\ref{fig:04_capacities_cf}), which is also reflected in the cost contribution (Figure~\ref{fig:06_cost_contribution_comparison}). This is due to the fact that capacity factors are significantly higher on average while investment costs per\,kW are comparable. Remarkably, the higher availability of solar power in Spain does not directly lead to higher PV capacities. In some years, optimum PV capacities are higher in the Netherlands than in Spain. Furthermore, the inter-annual variability is higher for PV capacities than for wind capacities – in contrast to the variability of the capacity factors. These surprising results are due to the fact that wind provides the main share of renewable power anyway. As a consequence timing and variability are decisive for the deployment of solar PV - not the overall generation. PV is useful only in times of low wind power abundance.

%

The different resources of wind speeds and solar radiation at the two regions lead to different operation strategies and different system layouts. The results show how difficult it is to design a storage solution which can be operated optimally for many years (i.e. fluctuating weather conditions). 

For 2016 the optimization process finds a battery storage system with 100\,kWh capacity as cost-optimal in Spain, in Netherlands only half the capacity is needed. Results for Netherlands show that in 14 years the battery storage is not used at all (the battery storage size is zero). Within the other 23 years the battery option is used, however its maximum size is at 90\,kWh, whereas in Spain the battery is used in every single year and its maximum size is at 243\,kWh.

\begin{figure}[tb]
    \centering
    \includegraphics[width=\columnwidth]{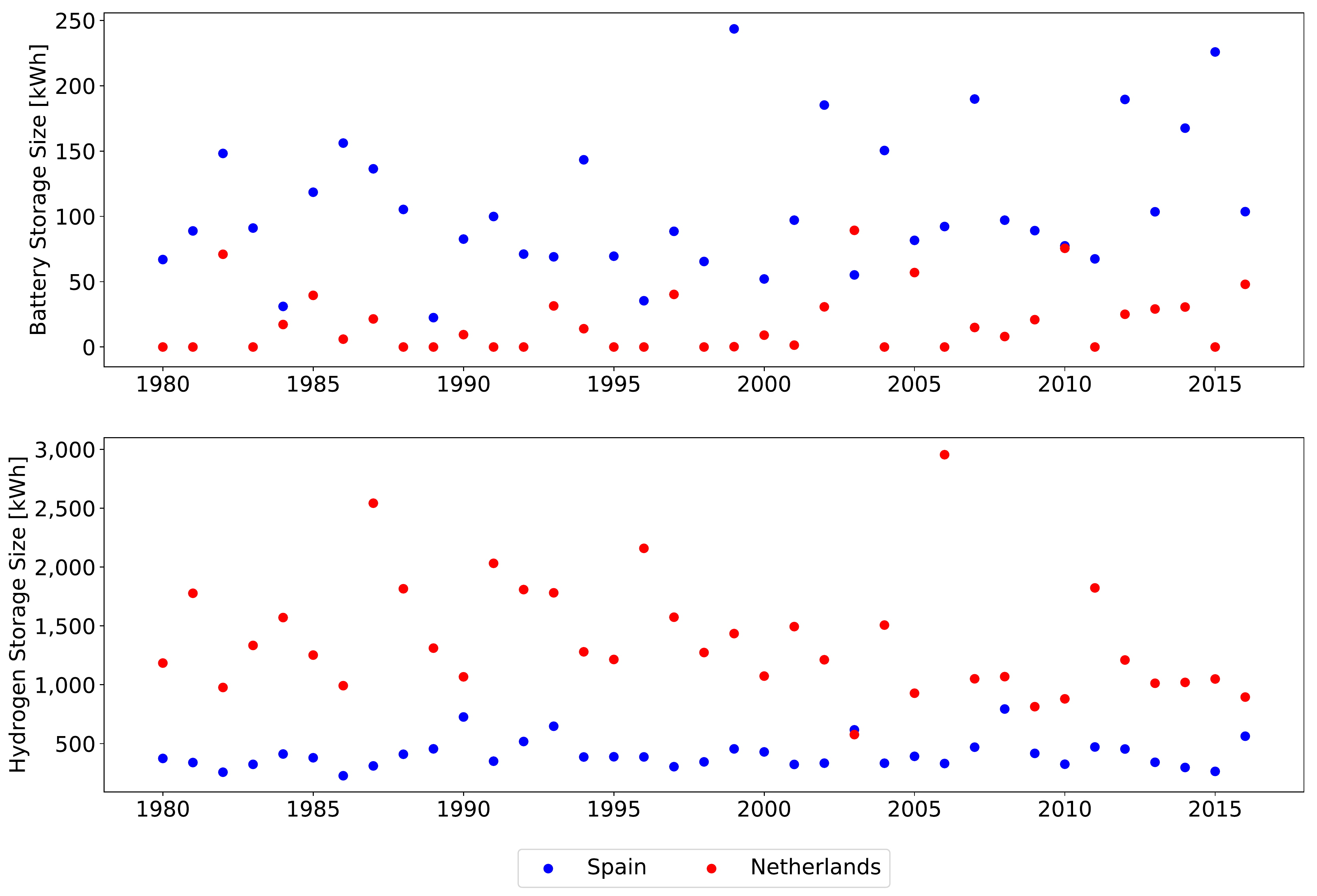}
    \caption{Comparison of storage sizes for all years.}
    \label{fig:05_storage_sizes}
\end{figure}

The hydrogen storage is larger in the Netherlands (900\,kWh corresponding to 27\,kg\,H$_{2}$) than it is in Spain (560\,kWh corresponding to 17\,kg\,H$_{2}$). Hydrogen storage is used in every model run, thus the option of an intermediate hydrogen storage is always cost beneficial. This tendency can be observed for almost all years (Figure~\ref{fig:05_storage_sizes}). Fluctuation of the size of the hydrogen storage, however, is very low in Spain compared to the Netherlands. The variance can be used for a good description of the fluctuation of variables, for the hydrogen storage the variance is 15 times higher for Netherlands than it is for Spain (Spain: 15,809\,kWh, Netherlands: 243,279\,kWh).

Often PV generation peaks cannot be consumed by the electrolyzer in Spain, as its input load is fixed to 100\,kW$_{el}$. A battery storage provides a short-term intermediate storage to consume generation peaks. In shortfalls the battery can provide the necessary electricity for hydrogen production to ensure the minimum load of the methanisation. Unlike in the Netherlands, where the intermediate hydrogen storage increasingly safeguards the minimum load. 

\subsection{Contribution of different technologies to the overall costs and room for technological improvements}
\label{sec:contribution-of-different-technologies-to-costs}

\begin{figure}[tb]
    \centering
    \includegraphics[width=\columnwidth]{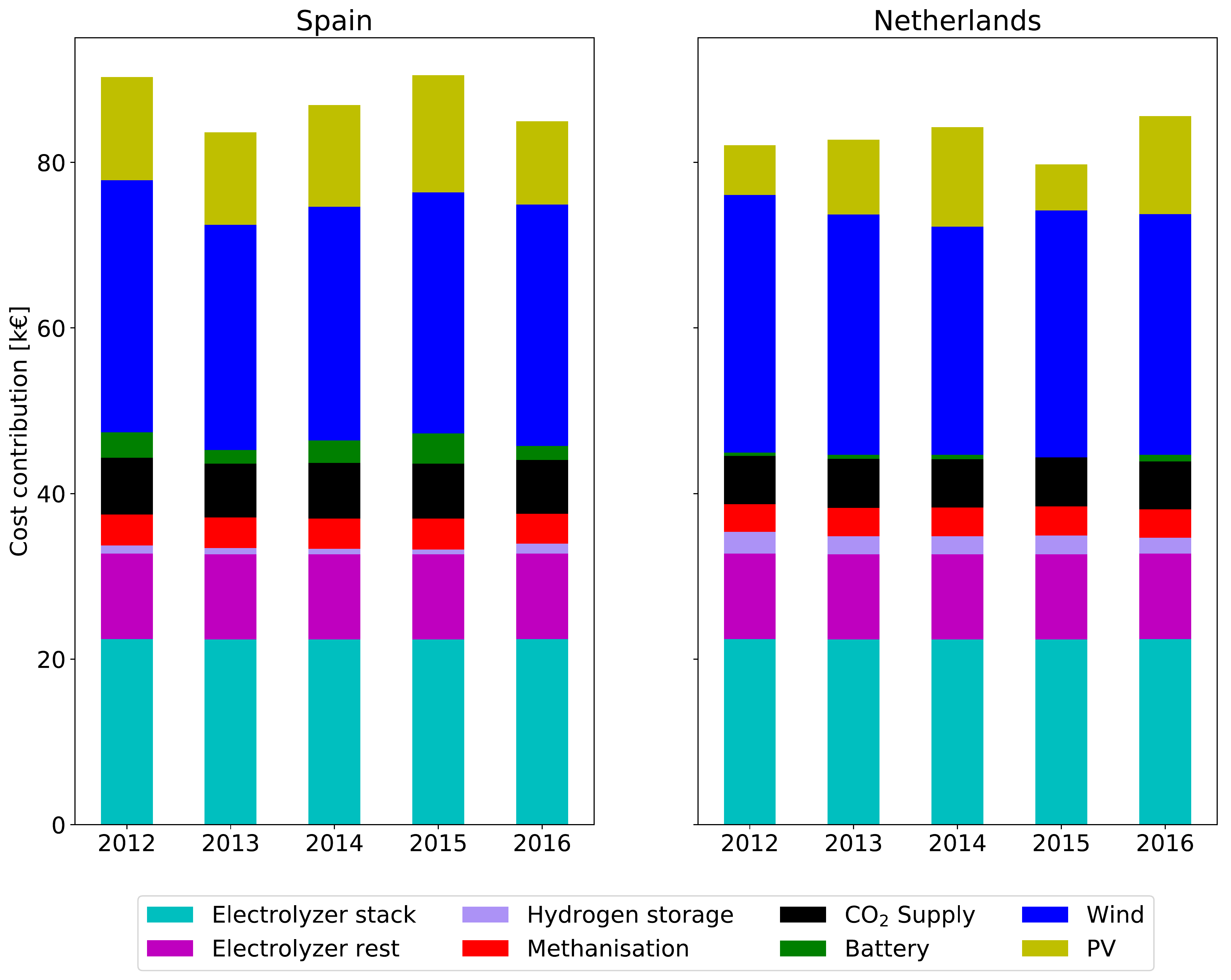}
    \caption{Comparison of cost contribution for five years (2012-2016).}
    \label{fig:06_cost_contribution_comparison}
\end{figure}

The absolute contribution of different technologies to the total system costs are summarized in Figure~\ref{fig:06_cost_contribution_comparison} for five years (2012 - 2016) and both regions. Different storage usage strategies are reflected in the costs, in Spain the contribution of the battery to the total system costs is higher than in the Netherlands and vice versa for the hydrogen storage. The cost contribution analysis (cf. Figure~\ref{fig:06_cost_contribution_comparison}) reveals that the highest share is accounted by the electricity supply technologies. It is slightly higher than those of the installation of the electrolyzer, where the influence of the stack components is dominant. Storage technologies, be it hydrogen storage or battery storage, do not contribute significantly to the total costs. Carbon dioxide supply and the methanisation reactor is a relevant factor for cost contribution. However, the total share is rather low. Thus, cost reductions for both are not able to reduce the overall system costs significantly.

\begin{figure}[tb]
    \centering
    \includegraphics[width=\columnwidth]{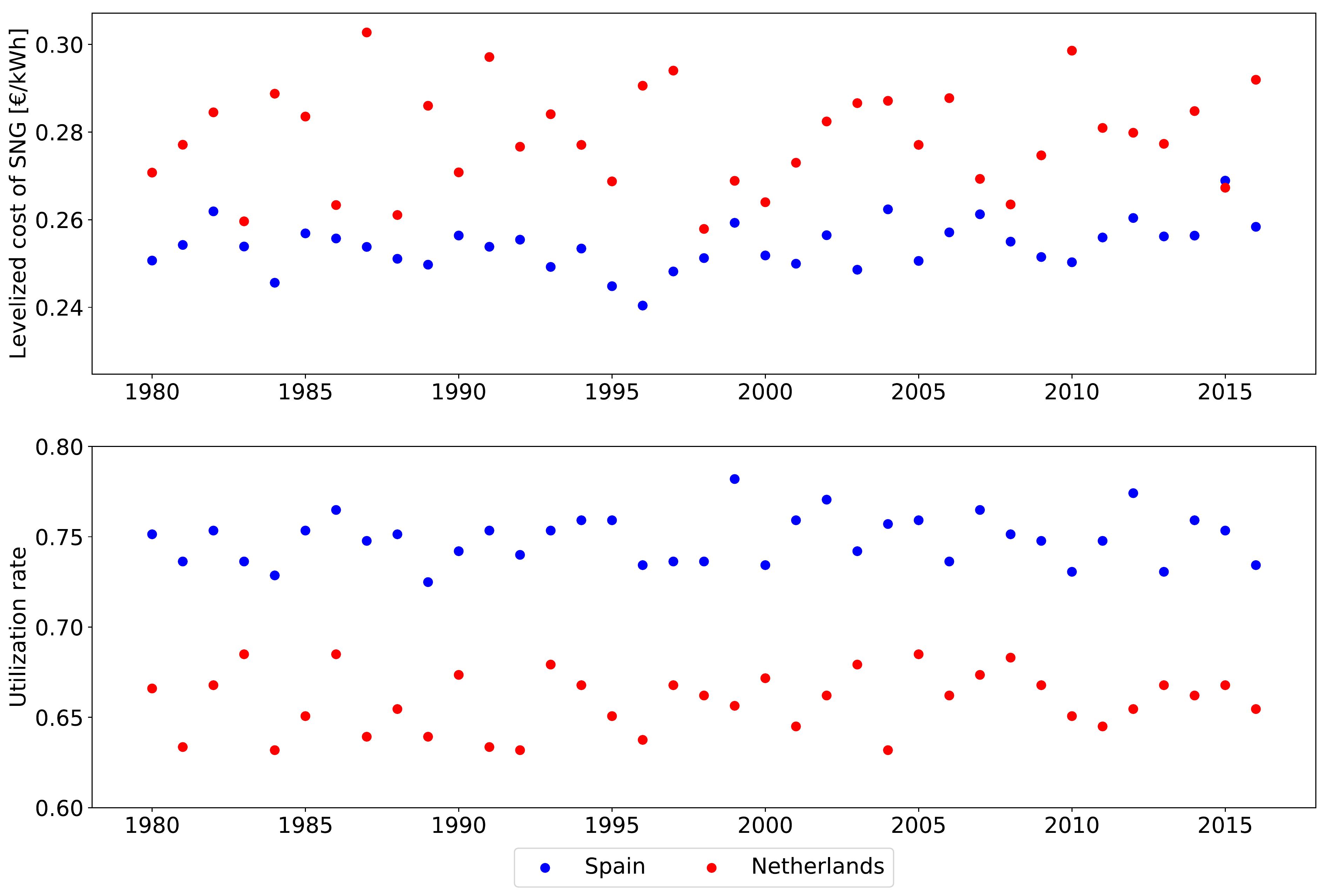}
    \caption{Levelized cost of SNG and the utilization rate of the PtM plants for both regions and all years.}
    \label{fig:07_levelized_cost_ur}
\end{figure}

The magnitude of the different total system costs varies from year to year. The cost-optimal full load hours is different between the years as well, thus the produced amount of SNG varies between the years. For a valid comparison, it is reasonable to calculate the levelized cost of SNG, which has been discussed in section~\ref{sec:methodological-background-to-levelized-cost-of-SNG}. Figure~\ref{fig:07_levelized_cost_ur} gives an overview of the levelized cost of SNG and the utilization rate of the PtM plant for all years and both regions. 

Spain shows better economic performance for almost all years, only in 2015 performance in Netherlands is slightly better (Spain: 26.89 Cent/kWh, Netherlands: 26.73 Cent/kWh). Generally, a higher utilization rate leads to a better economic performance.

\begin{figure}[tb]
    \centering
    \includegraphics[width=\columnwidth]{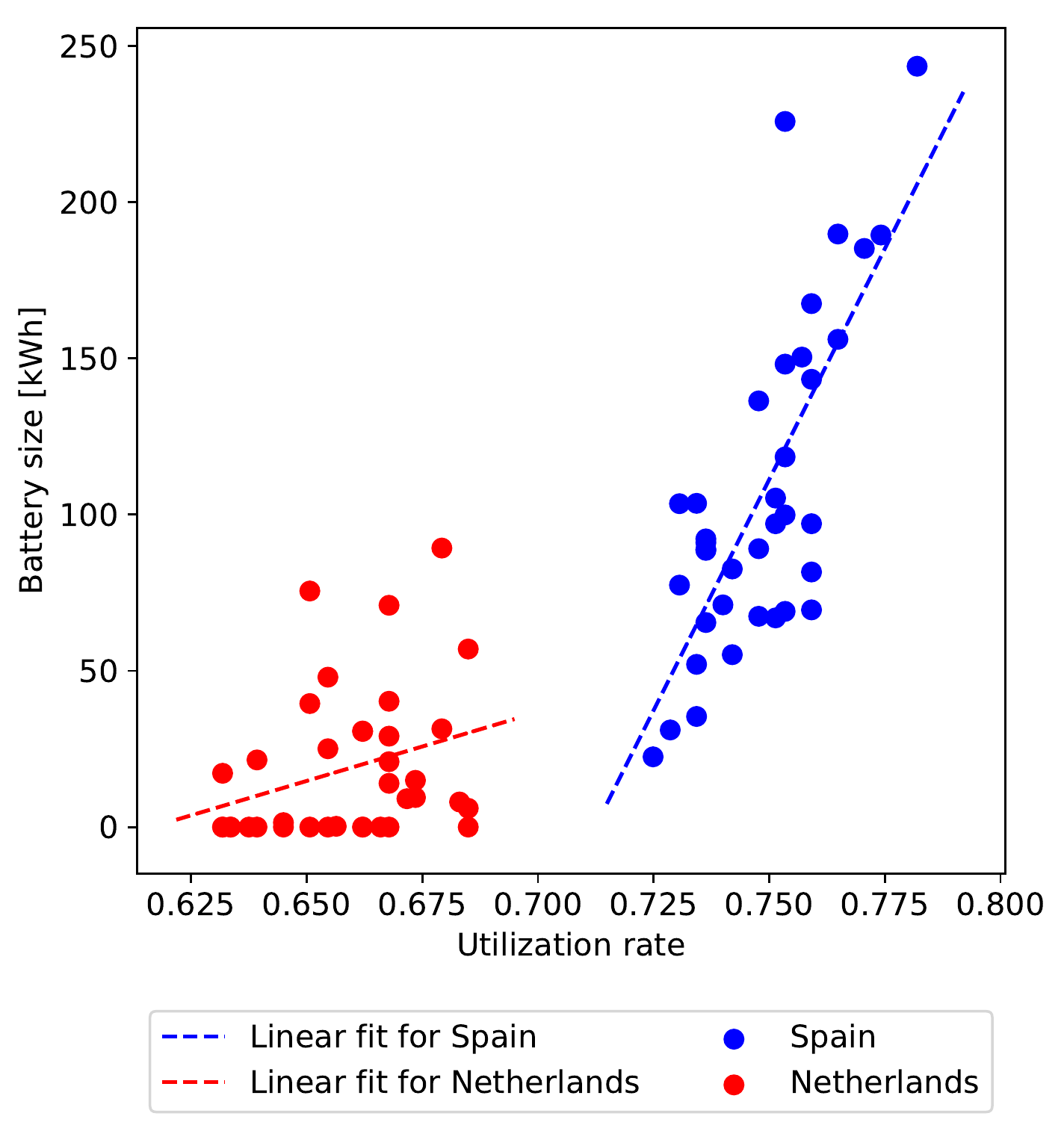}
    \caption{Comparison of battery size and utilization rate for Spain (blue) and the Netherlands (red). Circles correspond to different weather years and the dashed line is a linear fit to the data.}
    \label{fig:08_battery-size-over-ur}
\end{figure}

Technically, a larger battery storage enables a higher utilization rate of the electrolyzer, as it can still operate in times of low renewable generation. However, the effective relation of storage and utilization is more complex – it crucially depends on the temporal patterns of renewable generation and the investment costs of the battery. In fact, we find a strong correlation of storage size and utilization only for Spain (Figure~\ref{fig:08_battery-size-over-ur}, blue circles). In the Netherlands, the optimum battery size is typically smaller and in some cases not cost-beneficial and thus not used at all (Figure~\ref{fig:08_battery-size-over-ur}, red circles). In conclusion, a battery can be a crucial part of the island PtM plant by increasing its utilization rate, but this is not necessarily always the case. To further investigate the role of battery storage, we compare the presented model results with another optimization run which does not have the battery storage option in section~\ref{sec:importance-of-battery-storage}.

\subsection{Use of co-products: Oxygen and curtailed power}
\label{sec:use-of-coproducts}

\begin{figure}[tb]
    \centering
    \includegraphics[width=\columnwidth]{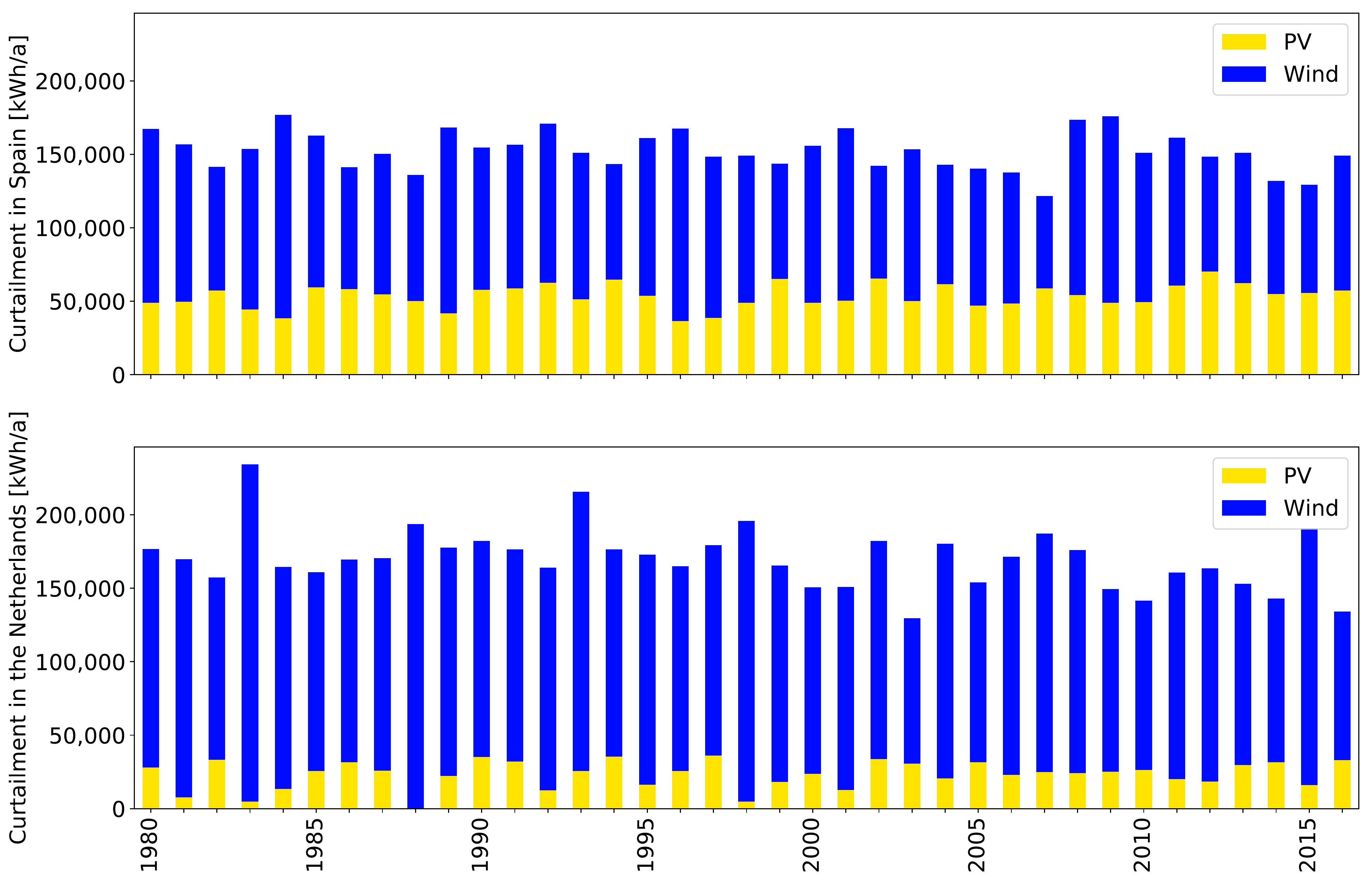}
    \caption{Curtailment of electricity from PV and wind for both regions and all years.}
    \label{fig:09_curtailment}
\end{figure}

Curtailment exists in all model runs, whenever the supply exceeds the demand. Figure~\ref{fig:09_curtailment} shows a comparison of the curtailment for Spain and the Netherlands. Generally curtailment in Spain is lower, however amounts are significant for both regions (around 150\,MWh/year). Let’s put this amount into perspective: If the electrolyzer with 100\,kW$_{el,in}$ would run the whole year at its full capacity it would require 876\,MWh of electricity per year.

Many use cases for the curtailed electricity are conceivable for practical applications: heat or steam generation being the simpler concepts. It remains questionable if the curtailed electricity has an economic value, which in principle would decrease the levelized cost of SNG. Different revenues from the sale of curtailed electricity lead to different levels of cost reduction. If the revenue of curtailed power is 1 Cent/kWh, this reduces costs by 1.8\,\%, whereas a revenue of 5 Cent/kWh can lead to cost reductions of around 9\,\%. This lever is rather large, thus use cases for the curtailed electricity can be crucial for the economic performance of the whole PtM plant. However, electricity is curtailed during periods of high renewable generation, i.e. at times where power generation is typically high such that market prices are low.

Similarly, oxygen accumulates as a co-product during the electrolysis and can potentially be utilized or sold for other applications. A reasonable price for oxygen depends heavily on amounts, operational parameters (such as pressure) and distance to the use cases. In the future, iron and steel plants might require higher amounts of oxygen to reduce their carbon dioxide footprint via the oxyfuel process, where oxygen is used instead of air. In this case the flue gas is carbon dioxide rich which can be stored (carbon capture and storage) or used (carbon capture and utilization) instead of releasing it to the atmosphere. The amount of produced oxygen is dependent on the utilization rate of the PtM plant and is approximately in the order of magnitude of 80t oxygen per year. Assuming an oxygen revenue of 1 Cent/kg leads to a cost reduction of approximately 1\,\%, however 10 Cent/kg lead to a cost reduction potential of approximately 10\,\%. 

\subsection{Importance of battery storage}
\label{sec:importance-of-battery-storage}
\begin{figure}[tb]
    \centering
    \includegraphics[width=\columnwidth]{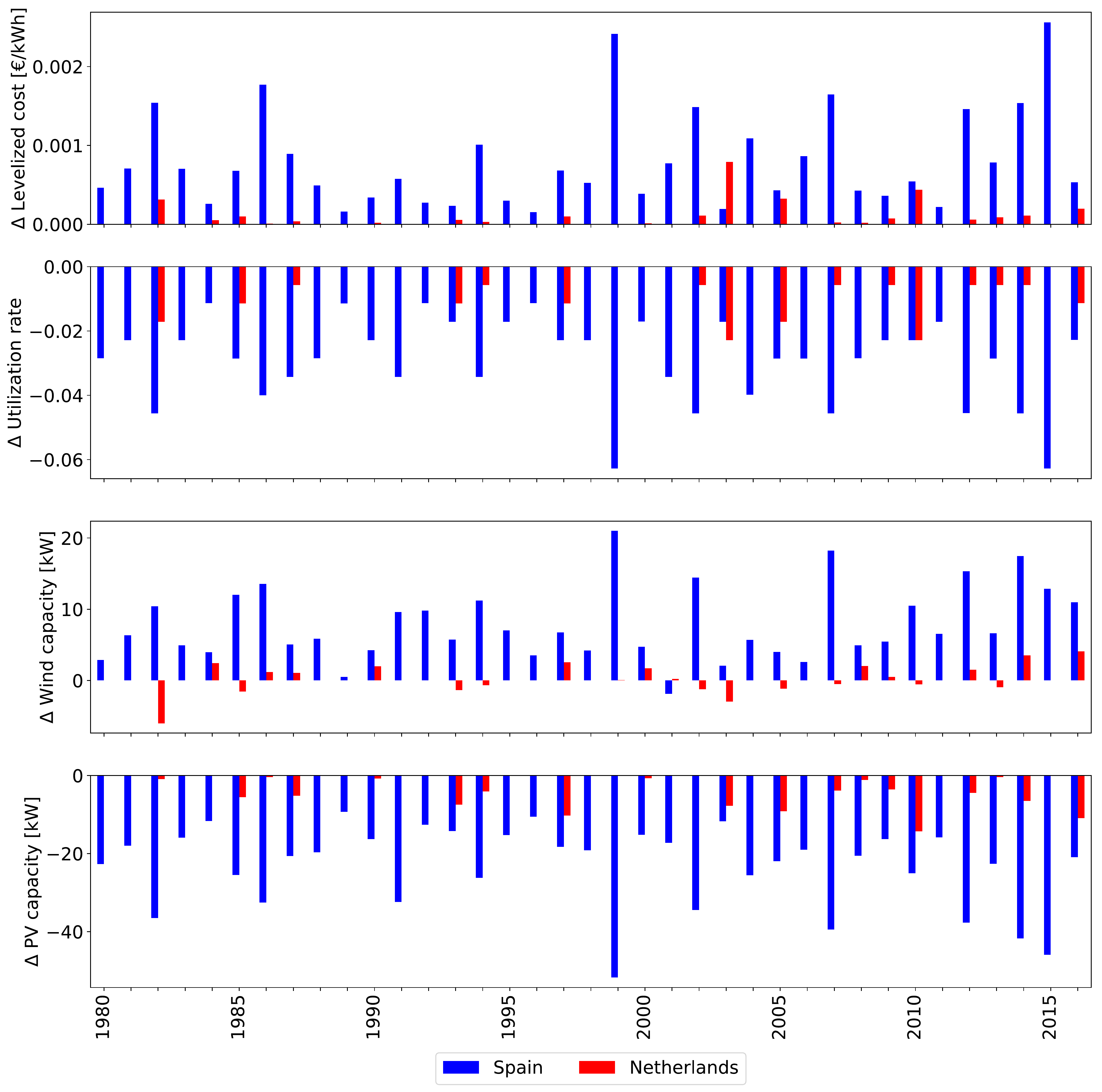}
    \caption{Difference between the battery and no battery cases.}
    \label{fig:10_delta_bat_vs_nobat}
\end{figure}

To quantify the importance of electricity storage, we compare the system optimum with and without a battery storage unit. In particular, we repeat all optimizations with the same parameters, but excluding the battery, and evaluate the difference of levelized costs, utilization rates and optimum renewable capacities with respect to the reference case (Figure~\ref{fig:10_delta_bat_vs_nobat}). Surprisingly, the abandonment of a battery has an almost negligible effect on the levelized costs of SNG. The increase of costs is below 0.3 Cent/kWh (below 1\,\%) in Spain for all years and even below 0.1 Cent/kWh (below 0.3\,\%) in the Netherlands. A stronger effect is observed in terms of the utilization rate of the electrolyzer, which drops by approx. 6\,\% or equivalently by 550h in certain years. A strong reduction in the utilization rate correlates with an increase of levelized costs – but the overall effect on levelized costs remains small as stressed above. Extensive effects are observed in terms of the optimum system layout for the region ES61 Spain. Abandoning the battery leads to a strong decrease of the optimum PV capacity, up to 50\,kW in certain years, which are compensated by an increase of the optimum wind power capacity. It can be assumed that because of the characteristics of PV generation it is beneficial to combine PV with a battery storage.

Due to the missing option of battery storage the amount of curtailment changes as well, however the changes are rather low. This is due to the oppositional trends for wind and PV capacities: The system generates more wind power, but less solar power and the effects even out to a large extent. In some cases the curtailment even decreases slightly, but in most cases it increases in an order of magnitude of approximately 10,000\,kWh/a for Spain and 2,000\,kWh/a for Netherlands.

The size of the hydrogen storage remains almost constant in the Netherlands for all years. In Spain it increases between 1.5\,kg and 10\,kg. The hydrogen storage is the only way to store energy and to ensure a constant load of the methanisation reactor above 40\,\%. Thus an increase of its size is needed in Spain, where the battery had a part of the important task to store energy for periods with no wind and PV generation. 

The results reveal two important properties of island PtM systems. Firstly, electricity storage is mainly used for intra-day load shifting of PV generation. This is also clearly visible in Figure~\ref{fig:load_balance}. As a consequence, excluding a battery reduces the PV utilization and capacity in the system optimum. Secondly, the electricity supply components of the PtM system are to a large extent exchangeable. A decrease in storage capacity leads to a shift from solar to wind power, but the levelized costs of SNG are hardly affected. Mathematically speaking, we observe an extremely flat minimum of the objective function. This leaves a lot of freedom in the design of the electricity subsystem of the PtM plant, such that secondary objectives can be taken into account in the design. In particular, electricity storage is not essential for island PtM plants, not even in Spain.

\subsection{Sensitivity analysis}
\label{sec:sensitivity-analysis}

\begin{figure}[tb]
    \centering
    \includegraphics[width=\columnwidth]{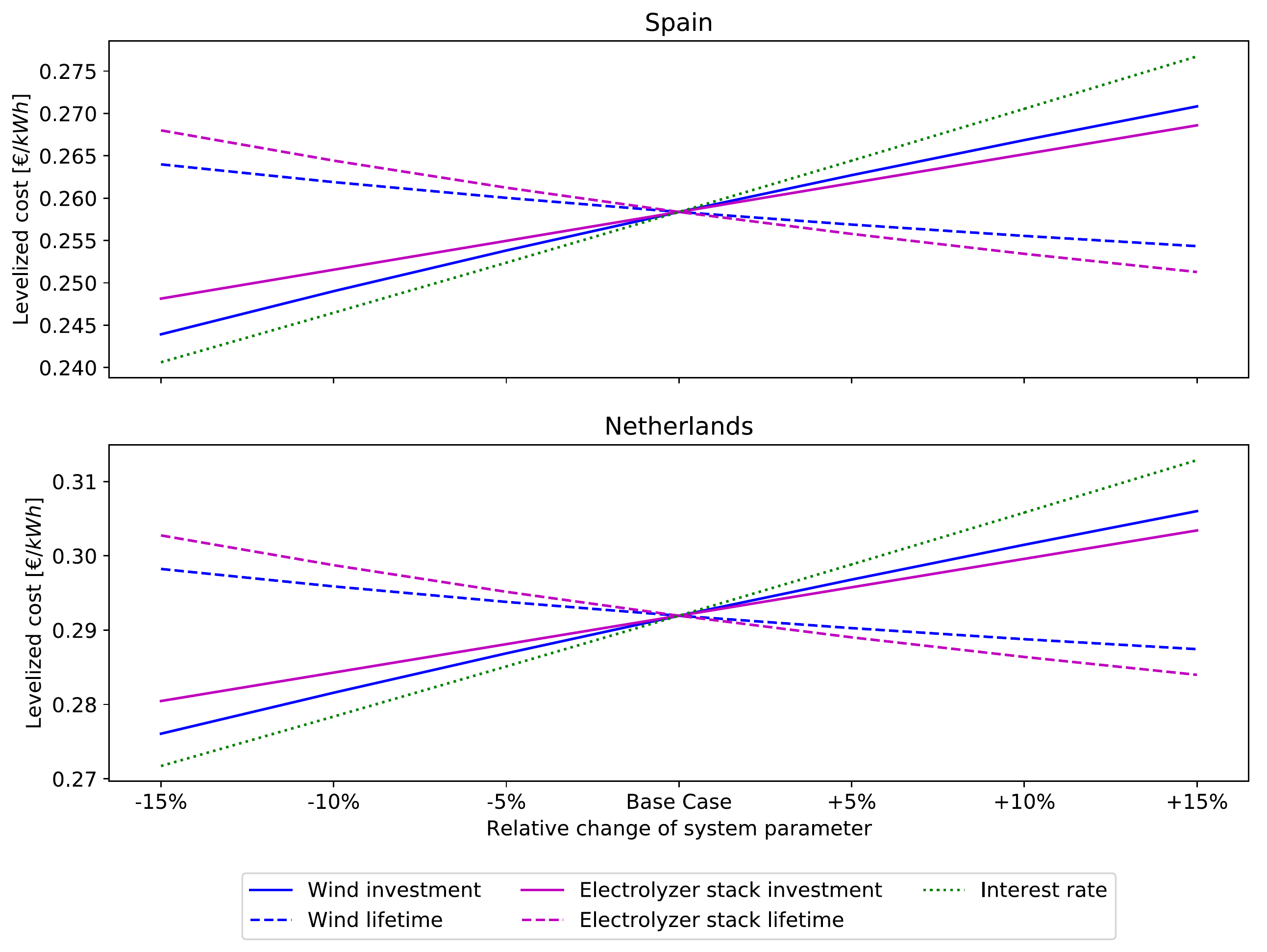}
    \caption{Sensitivity analysis of the levelized costs of SNG. We evaluate the change of the levelized costs when a single techno-economic parameter is changed, fixing all other parameters as well as the load level.}
    \label{fig:11_sensitivity}
\end{figure}

The costs of SNG crucially depend on the costs and lifetimes of system components and other techno-economic parameters (Table~\ref{tab:model-parameters}). We analyze the sensitivity of levelized costs of SNG to parameter changes to account for parameter uncertainties and to understand which technological developments can improve the economic competitiveness of PtM. For simplicity we focus on a single weather year (2016) and fix the load level at the optimum for the initial parameter settings (Netherlands: 5750h (=65.6\,\%), Spain: 6450h (=73.4\,\%)). Figure~\ref{fig:11_sensitivity} shows the results for the top five parameters with the highest impact on cost results. 

The impact of the parameters are very similar for both regions, with the highest impact being the interest rate, followed by wind investment, electrolyzer stack investment, electrolyzer stack lifetime and wind lifetime. Remarkably, a reduction of interest rates has the highest impact on the levelized costs. Hence, competitiveness of PtM depends rather strongly on the general economic framework, such as country specific WACCs. From the perspective of technological developments, a reduction of the investment costs for wind power, the dominant renewable energy source, has the highest impact. On the one hand, this is encouraging for PtM, since costs for renewable power generation has been declining strongly and are expected to continue declining. On the other hand, this result is discouraging for research on PtM system components, as the impact on the levelized costs is limited.

\section{Conclusion}
\label{sec:conclusion}

Power-to-Methane is a promising option to foster flexibility, sector coupling and the decarbonization of transport and heat in future energy systems. However, the economic competitiveness of SNG with respect to fossil alternatives remains an open issue. In this article we have optimized the layout and operation of island PtM systems at high spatio-temporal resolution. We have analyzed the necessary investments as well as resulting levelized costs and utilization rates with a focus on the roles of climatic conditions and the variability of renewable electricity sources.

We have examined in detail the need for flexibility options to deal with the variable power generation. Electricity is mainly consumed by the electrolyzer, which allows for a mostly flexible use. Hence, the main factor limiting its flexibility is not due to technical requirements, but due to economic considerations. A certain level of utilization is needed to amortize the investment costs. Remarkably, battery storage systems are of little importance for this task. Excluding this option from the model leads to only small changes in the levelized costs of SNG. In fact, one finds that the main use of a battery is the balancing of the solar daily profile. The most important flexibility options guaranteeing a high utilization level are an optimum mix of wind and solar power and a significant oversupply of renewable electricity, which is curtailed if necessary. In contrast, the methanisation reactor requires a certain minimum load for operation. Thus, a rather large hydrogen storage is required if the electrolyzer is operated in a flexible way. This aspect of PtM is of general importance also beyond the special case of an island system. Several technologies for sector coupling or CCU have high investment costs such that a high utilization is essential for their economic behavior. Hence, they must be operated as continuous as possible and cannot provide flexibility to the system. PtM is different – operation readily adapts to the availability of wind and solar power without an economic necessity for electricity storage and may thus provide flexibility. It has to be noted, however, that curtailment of electricity is an essential aspect of flexibility.

In an optimum system layout, electricity is mainly provided by wind power, while solar photovoltaics contributes between 0\,\% and 27\,\%. The inter-annual variability of wind power generation is rather large, in particular larger than the variability of solar power generation. As a consequence, also the optimum system layout varies from year to year and a compromise must be made when planning a real world system with a lifetime of several years. Remarkably, the optimum solar capacity varies much more than the optimum wind capacity. These findings admit two important conclusions. First, solar power is important only during times when wind power generation is low. Hence, the timing of solar power generation is more important than the total energy yield. Second, appropriate climatic data is needed for planning energy systems with a high renewable share – using just a few weather years can be misleading.
The system costs are mainly driven by electricity provision and the electrolyzer stack. Carbon dioxide supply, storage technologies and methanation reactor play minor roles. This can be seen in the sensitivity analysis, which reveals that the highest lever for cost reduction is the interest rate. This crucial parameter is dependent on many factors, varying depending on the industry sector in which the project is based and on the share of debt compared to equity in a project. For instance, the weighted average cost of capital in Germany was around 7\,\% in 2018 \cite{RN17}. The cost of equity is higher than the cost of debt \cite{RN16}, meaning that often large project developers, more able to secure debt financing, have an advantage over smaller developers.

Several by-products are produced during the PtM process, two of which have been discussed in detail in this paper. Both the curtailed electricity and the produced oxygen can be sold which can only improve the economic competitiveness of PtM. However, potential revenues are hard to quantify. For instance, electricity prices vary strongly with the renewable generation and future prices depend strongly on the renewable share and the availability of flexibility option in the grid. System simulations can be extended by taking into account by-products explicitly and evaluating different scenario for expected cost reductions.

The current work focused on the operation and optimization of a single PtM plant including electricity supply and storages. Future work should further elaborate on a detailed market analysis including regulations and financing costs and a discussion of potential technical developments. The implementation of (large-scale) PtM plants can lead to opportunities for different actors and countries. On a European level, SNG could be traded in the future.

\section*{Acknowledgments}
SM acknowledges funding by the German Federal Ministry of Education and Research (BMBF) within the Ko\-per\-ni\-kus Project P2X: Flexible use of renewable resources research, validation and implementation of Power-to-X concepts. JK acknowledges support via the center of excellence “Virtual Institute - Power to Gas and Heat” (EFRE-0400151) by the “Operational Program for the promotion of investments in growth and employment for North Rhine-Westphalia from the European fund for regional development” (OP EFRE NRW) through the Ministry of Economic Affairs, Innovation, Digitalization and Energy of the State of North Rhine-Westphalia. CB acknowledges funding by the Federal Ministry for Economic Affairs and Energy (BMWi). DW acknowledges support by the Helmholtz Association (via the joint initiative “Energy System 2050 - A Contribution of the Research Field Energy'' and the grant no. VH-NG-1025).

\section*{Bibliography}


\end{document}